# The Maxwell wave function of the photon

M. G. Raymer and Brian J. Smith

Oregon Center for Optics and Department of Physics

University of Oregon, Eugene, Oregon 97403

## ABSTRACT

James Clerk Maxwell unknowingly discovered a correct relativistic, quantum theory for the light quantum, forty-three years before Einstein postulated the photon's existence. In this theory, the usual Maxwell field is the quantum wave function for a single photon. When the non-operator Maxwell field of a single photon is second quantized, the standard Dirac theory of quantum optics is obtained. Recently, quantum-state tomography has been applied to experimentally determine photon wave functions.

> "But to determine more absolutely what light is, after what manner refracted, & by what modes or actions it produceth in our minds the Phantasms of colours, is not so easie. And I shall not mingle conjectures with certaintyes." [+]
>
> – Isaac Newton

## 1. THE MAXWELL PHOTON WAVE FUNCTION

In about 1862, James Clerk Maxwell determined mathematically from his then-new equations, that electromagnetic waves travel at a speed very nearly equal to the measured value of the speed of light. In 1864 he wrote [1],

> "This velocity is so nearly that of light that it seems we have strong reason to conclude that light itself (including radiant heat and other radiations) is an electromagnetic disturbance in the form of waves propagated through the electromagnetic field according to electromagnetic laws."

In 1862 he wrote in *On Physical Lines of Force* [1],

> "We can scarcely avoid the inference that light consists in the traverse undulations of the same medium which is the cause of electric and magnetic phenomena."

Maxwell's equations are, for a source-free region of space (in Gaussian units),

$$\frac{\partial}{\partial t}\vec{E}(\vec{r},t) = c\vec{\nabla}\times\vec{B}(\vec{r},t), \quad \frac{\partial}{\partial t}\vec{B}(\vec{r},t) = -c\vec{\nabla}\times\vec{E}(\vec{r},t)$$
$$\vec{\nabla}\cdot\vec{E}(\vec{r},t) = 0, \quad \vec{\nabla}\cdot\vec{B}(\vec{r},t) = 0 \qquad (1)$$

Max Planck said [1], on the centenary of Maxwell's birth in 1931, that Maxwell's theory "...remains for all time one of the greatest triumphs of human intellectual endeavor."

---

[+] *A Theory Concerning Light and Colors*, Cambridge University Library Add MS 3970.3 ff. 460-66, http://www.newtonproject.ic.ac.uk/texts/cul3970_n.html

Planck was correct—even more so than he realized. For, just a year earlier, in 1930, Paul Dirac had shown the way to formulate dynamical equations for relativistic elementary particles. It is now understood that Dirac's particle approach, when applied to massless spin-one particles, leads directly to Maxwell's equations. This means that Maxwell unknowingly discovered a correct relativistic, quantum theory for the light quantum, forty-three years before Einstein postulated the photon's existence! In this theory, the (non-operator) Maxwell field is the quantum wave function for a single photon. When the non-operator Maxwell field of a single photon is quantized, the standard Dirac theory of quantum optics is obtained.

Here we review the derivation of Maxwell's equations from relativistic, quantum particle dynamics, which in recent times was expounded on in detail by Bialynicki-Birula [2] and by Sipe [3], and later by Kobe [4]. We follow [2] and [3], while trying to present a simpler version of the derivation.

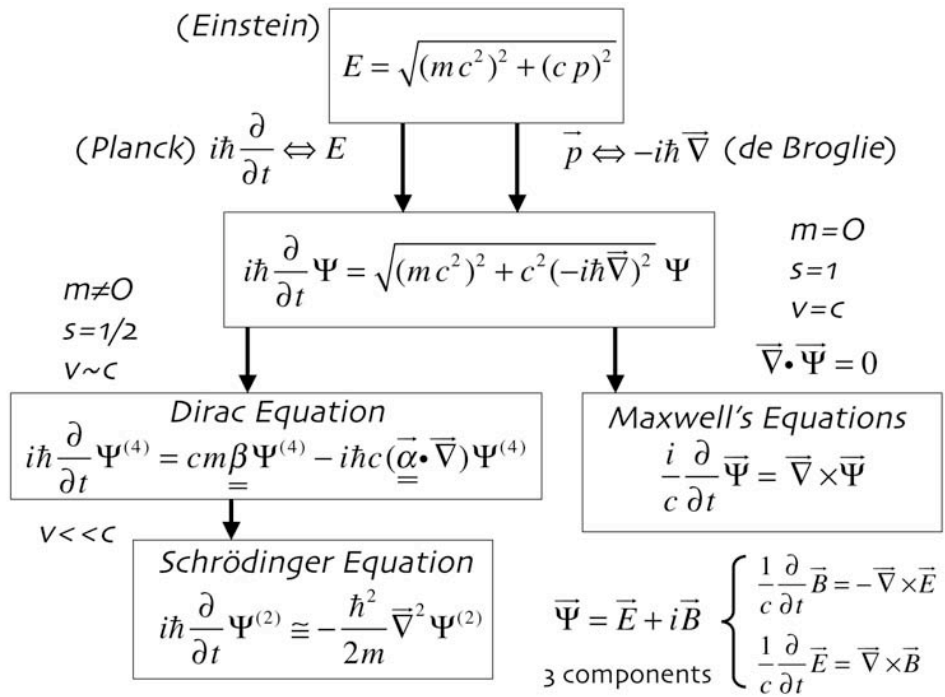

Fig.1 Flow chart for derivations of electron and photon wave equations, $m$ = rest mass, $s$ = spin, $v$ = velocity.

In modern terms, a photon is an elementary excitation of the quantized electromagnetic field. If it is known a priori that only one such excitation exists, it can be treated as a (quasi-) particle, roughly analogous to an electron. It has unique properties, arising from its zero rest mass and its spin-one nature. In particular, there is no position operator for a photon, leading some to conclude that there can be no properly defined wave function, in the Schroedinger sense, which allows localizing the particle to a point. On the other hand, it is known that even electrons, when relativistic, don't have properly defined wave functions in the Schroedinger sense [2,3], and this opens our minds to broader definitions of wave functions. In relativistic quantum theory, one distinguishes between charge-density amplitudes, mass-density amplitudes, and particle-number density amplitudes. These can have different localization properties. Photons, of course, are inherently relativistic, so it is not surprising that we need to be careful about defining their wave functions.

Dirac's theory of a particle is based on the kinematic equation for energy $E$, momentum $\bar{p} = (p_x, p_y, p_z)$, and rest mass, $m$,

$$E = \sqrt{(mc^2)^2 + c^2 \vec{p} \cdot \vec{p}} \quad . \tag{2}$$

Define a multicomponent amplitude function $\widetilde{\psi}(\vec{p}, E)$ obeying the normalization condition

$$(2\pi\hbar)^{-3} \int d^3p \, \widetilde{\psi}^*(\vec{p}, E) \cdot \widetilde{\psi}(\vec{p}, E) = 1, \tag{3}$$

where the dot indicates a vector dot product. Multiplying Eq.(2) by this function gives

$$E\widetilde{\psi}(\vec{p}, E) = \sqrt{(mc^2)^2 + c^2 \vec{p} \cdot \vec{p}} \, \widetilde{\psi}(\vec{p}, E). \tag{4}$$

For a spin-one-half particle with non-zero rest mass, this equation gives, upon recognizing that the wave function $\widetilde{\psi}(\vec{p}, E)$ must have two components (for the positive-energy solutions), and transforming to space-time variables $\vec{r}, t$, the Dirac equation for the electron. The electron wave function is a two-spinor, $\psi(\vec{r}, t) = (\psi_{1/2}, \psi_{-1/2})$, where the components $\psi_{\pm 1/2}$ are amplitudes for the states of plus- and minus-1/2 spin-projection onto the quantization axis. In this case, Eq.(3) represents normalization of the probability to find particular values of the electron's momentum.

Because a photon is massless, its wave function should obey

$$E\widetilde{\psi}(\vec{p}, E) = c\sqrt{\vec{p} \cdot \vec{p}} \, \widetilde{\psi}(\vec{p}, E), \tag{5}$$

Since the photon is a spin-one particle, its wave function should have three components, forming a (non-operator) three-component vector field $\widetilde{\psi}(\vec{p}, E) = (\widetilde{\psi}_x, \widetilde{\psi}_y, \widetilde{\psi}_z)$. To represent the square-root operator $\sqrt{\vec{p} \cdot \vec{p}}$ we look for a vector operator $\hat{A}$ with the property $(\hat{A})^2 \widetilde{\psi} = \vec{p} \cdot \vec{p} \, \widetilde{\psi}$. Such an operator can be found by elementary means, by trying $\hat{A} = i\vec{p} \times$, where $\times$ is the cross-product operator. Then a well-known vector identity gives

$$\hat{A}\hat{A}\widetilde{\psi} = -\vec{p} \times (\vec{p} \times \widetilde{\psi}) = (\vec{p} \cdot \vec{p})\widetilde{\psi} - \vec{p}(\vec{p} \cdot \widetilde{\psi}). \tag{6}$$

Any vector field can be written as the sum of two linearly independent parts, $\widetilde{\psi} = \widetilde{\psi}_T + \widetilde{\psi}_L$, where the transverse part obeys $\vec{p} \cdot \widetilde{\psi}_T = 0$, and the longitudinal part obeys $\vec{p} \times \widetilde{\psi}_L = 0$. Identifying the transverse part as the relevant field for the photon, we derive the equivalent of Eq.(5),

$$E\widetilde{\psi}_T(\vec{p}, E) = ci\vec{p} \times \widetilde{\psi}_T(\vec{p}, E). \tag{7}$$

This deceptively simple-looking equation is actually equivalent to Maxwells' equations. To see this, first note that $\widetilde{\psi}_T$ must be a complex-valued vector if Eq.(7) is to be satisfied. Next, Fourier transform the amplitude function $\widetilde{\psi}(\vec{p}, E)$ from momentum space to coordinate space, and from energy to time, accounting for the constraint between energy and momentum ($E = c|\vec{p}|$) by including a delta function. This allows $E$ to be considered as an independent variable, and gives

$$\vec{\psi}(\vec{r}, t) = (2\pi\hbar)^{-4} \iint dE \, d^3p \, \delta(E - c|\vec{p}|) \exp(-iEt/\hbar + i\vec{p} \cdot \vec{r}/\hbar) f(E) \widetilde{\psi}(\vec{p}, E). \tag{8}$$

The momentum-space weight function $f(E)$ has been included to allow different forms of normalization of the coordinate-space function $\overline{\psi}_T(\vec{r},t)$. (In the case of the electron, discussed above, the standard choice is $f(E)=1$.) For the photon, we adopt the choice advocated by Sipe [3], $f(E) = \sqrt{E}$, which gives for the coordinate-space normalization,

$$\int d^3r\, \overline{\psi}(\vec{r},t)^* \cdot \overline{\psi}(\vec{r},t) = (2\pi\hbar)^{-3} \int d^3p\, E(p)\widetilde{\psi}^*(\vec{p},E(p)) \cdot \widetilde{\psi}(\vec{p},E(p)) = \langle E \rangle. \qquad (9)$$

where we defined $E(p) = c|\vec{p}|$, and $\langle E \rangle$ denotes the expectation value of the photon's energy. This choice of normalization reflects the fact that a photon has no mass that can be localized at a point; rather it has only helicity and energy, and the energy cannot strictly be localized at a point. The function $\left(\overline{\psi}(\vec{r},t)^* \cdot \overline{\psi}(\vec{r},t)\right)/\langle E \rangle$ is the probability density for energy, not particle location [2,3].

Equations (7) and (8) together give the "complex Maxwell equations,"

$$i\frac{\partial}{\partial t}\overline{\psi}_T(\vec{r},t) = c\vec{\nabla} \times \overline{\psi}_T(\vec{r},t). \qquad (10)$$

Notice that $\hbar$ acts only as a scaling factor in the Fourier transform functions, and cancels in Eq.(10). Also note that we did not have to postulate the de Broglie relation, $\vec{p} = -i\hbar\vec{\nabla}$; rather it emerges naturally from the Fourier transform. Further note that he transverse part of the field defined in Eq.(8) has zero divergence, $\vec{\nabla} \cdot \overline{\psi}_T = 0$, and the longitudinal part has zero curl, $\vec{\nabla} \times \overline{\psi}_L = 0$.

Now write the complex wave function as a sum of real and imaginary parts $\vec{E}_T(\vec{r})$ and $\vec{B}_T(\vec{r})$,

$$\overline{\psi}_T(\vec{r},t) = 2^{-1/2}\left(\vec{E}_T(\vec{r},t) + i\vec{B}_T(\vec{r},t)\right). \qquad (11)$$

Using Eq.(10), the real and imaginary parts $\vec{E}_T(\vec{r})$ and $\vec{B}_T(\vec{r})$ are found to obey Maxwell's equations, Eq.(1). Therefore, to paraphrase Maxwell's quote above, we can scarcely avoid the inference that the photon's quantum wave function consists in the traverse undulations of the same medium which is the cause of electric and magnetic phenomena. That is, the classical Maxwell equations are the wave equation for the quantum wave function $\overline{\psi}_T$ of a photon. Evidently, the longitudinal part of the $\overline{\psi}$ function corresponds to longitudinal electric and magnetic fields, which are non-propagating.

As a check, calculate the space normalization integral,

$$\int d^3r\, \overline{\psi}(\vec{r},t)^* \cdot \overline{\psi}(\vec{r},t) = \int d^3r\, \tfrac{1}{2}\left(\vec{E}_T \cdot \vec{E}_T + \vec{B}_T \cdot \vec{B}_T\right) = \langle E \rangle \qquad , \qquad (12)$$

which has the proper meaning that $\tfrac{1}{2}\left(\vec{E}_T \cdot \vec{E}_T + \vec{B}_T \cdot \vec{B}_T\right)$ is the local energy density.

The above derivation is for a particular helicity (handedness) of the photon angular momentum. The opposite helicity is described by changing Eq.(11) to $\overline{\psi}_T(\vec{r},t) = 2^{-1/2}\left(\vec{E}_T(\vec{r},t) - i\vec{B}_T(\vec{r},t)\right)$, and multiplying the right-hand side of Eq.(10) by $-1$.

## 2. MEASURING THE MAXWELL PHOTON WAVE FUNCTION

If a single-photon state of the field is created, then to know its quantum state means to know its electric and magnetic field distributions in space and time. Such a state is a single-photon wave-packet state, and its generation is an important goal in quantum-information research.

Recently, a technique has been developed to measure the transverse spatial quantum state of an ensemble of identically prepared photons [5, 6]. The single-photon light beam is sent into an all-reflecting, out-of-plane Sagnac interferometer, which performs a relative rotation of 180° and a mirror inversion on the wave fronts of the counter-propagating beams. The Sagnac performs a two-dimensional parity operation on one of the beams relative to the other. The fields are recombined at the output beam splitter and are interfered on a photon-counting photomultiplier tube (PMT), allowing the emerging beams to be detected at the single-photon level. The mean photo-count rate is directly proportional to the transverse spatial Wigner function at a phase-space point that is set by the tilt and translation of a mirror external to the interferometer.

The situation becomes even more interesting when the joint spatial wave function of a pair of photons is considered. In the case that the two photons' spatial and momentum variables are described by an entangled state, such a state measurement will provide the maximal-information characterization of the entanglement. By sending two entangled photons into two parity-inverting interferometers, one can measure the joint two-photon transverse spatial Wigner function, and completely characterize the transverse entanglement of this system [5,6]. The two-photon wave function exists in six spatial dimensions, and its equation of motion can be called the two-photon Maxwell's equations.

To conclude, the usual (classical) Maxwell field is the quantum wave function for a single photon. That it transforms like a three-dimensional vector arises from the spin-one nature of the photon. (In contrast, the electron transforms like a two-dimensional spinor.) When two photons are present, the joint wave function "lives" in a higher dimensional space. These observations imply the interpretation of the Maxwell field as akin to the Schroedinger wave function, which evolves probability amplitudes for various possible quantum events in which the electron's position is found to be within a certain volume, rather than being a realistic description of the electron as being here or there. In this sense, the Maxwell equation evolves the probability amplitudes for various possible quantum events in which the photon's energy is found within a certain volume. In addition, quantum-state tomography methods have been devised for determining spatial states of one- and two-photon fields.

We thank Cody Leary for helpful discussions about field theory. This research was supported by the National Science Foundation's ITR program, grant no. 0219460. M. G. Raymer's e-mail address is raymer@uoregon.edu.

1. From the web site Light through the ages: Relativity and quantum era, http://www-groups.dcs.st-and.ac.uk/~history/HistTopics/Light_2.html
2. I. Bialynicki-Birula, Acta Phys. Pol. **34**, 845 (1995); and "Photon wave function," in Progress in Optics XXXVI, E. Wolf, ed. (Elsevier, Amsterdam, 1996); and Phys. Rev. Lett. **80**, 5247 (1998).
3. J. E. Sipe, Phys. Rev. A **52**, 1875-1883 (1995).
4. D. H. Kobe, Found. Phys, **29**, 1203 (1999).
5. E. Mukamel, K. Banaszek, I. A. Walmsley and C. Dorrer, Opt. Lett. **28**, 1317-1319 (2003).
6. Brian J. Smith, Bryan Killett, Andrew Nahlik, M. G. Raymer, K. Banaszek and I. A. Walmsley "Measurement of the transverse spatial quantum state of light at the single-photon level," submitted for publication (Optics Letters, 2005).